\documentclass[epj]{svjour}
\pdfoutput=1
\usepackage{dcolumn}
\usepackage{amsmath,amssymb}
\usepackage{bm}
\usepackage{graphicx}
\newcommand{\bd}{\bm}

\begin{document}

\title{Bose-Einstein condensation at finite momentum and magnon condensation in thin film ferromagnets}

\author{J. Hick, F. Sauli, A. Kreisel, and P. Kopietz}
  
\institute{Institut f\"{u}r Theoretische Physik, Universit\"{a}t
  Frankfurt,  Max-von-Laue Strasse 1, 60438 Frankfurt, Germany}

 \date{October 22, 2010}

 \abstract{
We use the Gross-Pitaevskii equation to determine the spatial 
structure of the condensate density of interacting bosons
whose energy dispersion $\epsilon_{\bd{k}}$ 
has two degenerate minima at finite
wave-vectors  $ \pm \bd{q}$. 
We show that
in general the Fourier transform of the condensate density has  finite
amplitudes for all integer multiples of $\bd{q}$.
If the interaction is such that many Fourier components contribute,
the Bose condensate is  localized
at the sites of a one-dimensional lattice with 
spacing $2 \pi / | \bd{q} |$; in this case Bose-Einstein condensation
resembles the transition from a liquid to a crystalline solid. 
We use our results to investigate the spatial structure of the
Bose condensate
formed by magnons in thin films of ferromagnets with dipole-dipole interactions.
}
\authorrunning{J. Hick, et al.}
\titlerunning{BEC at finite momentum and magnon condensation in thin film ferromagnets}
\maketitle

\section{Introduction}

This work is motivated by the recent discovery~\cite{Demokritov06,Demidov07,Dzyapko07,Demidov08,Demokritov08}
of a new coherence phenomenon 
of magnons in thin stripes made of the magnetic insulator
yttrium-iron garnet (YIG). The energy dispersion
$\epsilon_{\bd{k}}$  of the lowest magnon mode
in this system has a rather unusual momentum-dependence which is crucial
to understand the experiments:
for  a certain range of orientations of
an external magnetic field, $\epsilon_{\bd{k}}$  exhibits
two degenerate minima at finite wave-vectors $+ \bd{q}$ and $- \bd{q}$.
The value of $\bd{q}$ is determined by a subtle interplay between
exchange interactions, dipole-dipole interactions, and 
finite-size effects~\cite{Kalinikos86,Kreisel09}.
The experimentally observed strong
enhancement of the occupation of the magnon modes with wave-vectors $\pm \bd{q}$
has been interpreted as Bose-Einstein condensation (BEC) of magnons.

Another class of boson systems where the energy dispersion has degenerate minima
at finite wave-vectors $\pm {\bd{q}}$ are magnon gases in  quantum helimagnets
in a magnetic field~\cite{Ueda09}.
Apart from the experimental relevance, the general  problem of BEC 
in systems where the energy dispersion  has a minimum 
for non-zero wave-vectors is  interesting on its own.
In this context we mention the work by
Yukalov~\cite{Yukalov78},  who has investigated BEC in an interacting Bose system 
whose energy dispersion is minimal on a sphere in momentum space. 
He found that in this case
the condensed state 
neither exhibits off-diagonal long-range nor is it superfluid.
Moreover, Yukalov also pointed out
an interesting analogy between BEC at finite momentum and
the liquid-crystal phase transition, which can also be understood in terms
of a Ginzburg-Landau functional whose Gaussian term
exhibits  minima on a surface in momentum space~\cite{Alexander78,Anderson84}.

The fact that BEC of quasi-particles is not necessarily accompanied by superfluidity
has been emphasized by Kohn and Sherrington \cite{Kohn70},
who classified bosons into two different
types: the first type consists of bound complexes of
an even number of fermions; in the case of condensation of these bosons, superfluidity 
and off diagonal long-range order occurs.
The second type of bosons consists of quasi-particles such as excitons and magnons;
when the second type condenses, there is no superfluidity, but a change
of spatial or magnetic order~\cite{Kohn70,superfluidity}.
Obviously, BEC of magnons in YIG is of the second type. 
From the point of view of critical phenomena it is not surprising
that BEC at finite momentum is rather different from BEC at zero momentum. In fact,
phase transitions which are characterized by an order
parameter which  condenses on a surface in momentum space
belong to their own universality class, the so-called
Brazovskii universality class~\cite{Brazovskii75,Hohenberg95}.

In this work we shall examine the general problem of BEC 
in a Bose gas whose energy dispersion has degenerate minima at
two finite wave-vectors $\pm \bd{q}$.
We show that in this case
the time-independent Gross-Pitaevskii equation
implies that the Fourier transform $\phi_{\bd{k}}$ of the
condensate wave-function has finite amplitudes $\phi_{0}, \phi_{ \pm \bd{q}},
\phi_{ \pm 2 \bd{q}}, \ldots $
for integer multiples of the fundamental wave-vector $\bd{q}$.
Previously a theoretical analysis of magnon-BEC in YIG
has been performed by Tupitsyn, Stamp, and Burin~\cite{Tupitsyn08}. 
However, the effect of the
spatial structure of the condensate
wave-function as implied by the Gross-Pitaevskii equation was not considered 
by these authors.

\section{BEC at finite momentum}

In this section we shall study BEC in a general class of interacting boson models on a lattice
whose Hamiltonian is of the form
 \begin{equation}
 H = H_2 + H_3 + H_4.
 \label{eq:Hdef}
 \end{equation}
The quadratic part $H_2$ of the Hamiltonian is given by
 \begin{equation}
 H_2 = \sum_{\bd{k}} 
 \left[ \epsilon_{\bd{k}} a^{\dagger}_{\bd{k}} a_{\bd{k}}
 + \frac{\gamma_{\bd{k}}}{2} a^{\dagger}_{\bd{k}} a^{\dagger}_{ - \bd{k}} +
\frac{\gamma^{\ast}_{\bd{k}}}{2} a_{-\bd{k}} a_{  \bd{k}}  \right],
 \label{eq:H2def}
 \end{equation}
where $a_{\bd{k}}$ and $a^{\dagger}_{\bd{k}}$ are the usual canonical annihilation and
creation operators, 
the energy dispersion $\epsilon_{\bd{k}}$ is assumed to exhibit
two degenerate minima at finite wave-vectors $\pm \bd{q}$, and the terms proportional to
the complex parameter $\gamma_{\bd{k}}$ explicitly break the $U(1)$ symmetry 
associated with particle number conservation. In YIG these terms are related to an external pumping
filed, as explained in the appendix.
In the absence of $U(1)$ symmetry, the Hamiltonian
can also contain  contributions involving three powers of 
boson operators, which in general are of the form
 \begin{eqnarray}
 H_3 & = & \frac{1}{\sqrt{N}} \sum_{ \bd{k}_1 \bd{k}_2 \bd{k}_3 } 
 \delta_{ \bd{k}_1 + \bd{k}_2 + \bd{k}_3 ,0}
 \Bigr[ 
 \nonumber
 \\
 & &
\frac{1}{2} \Gamma^{ \bar{a} aa}_{ 1; 2 3 }   
a^{\dagger}_{ -1} a_{ 2 } a_{3 } 
 +  \frac{1}{2} \Gamma^{ \bar{a} \bar{a}a}_{ 1 2 ; 3 }   
a^{\dagger}_{ -1} a^{\dagger}_{ -2 } a_{3 } 
 \nonumber
 \\
 &+ &
  \frac{1}{3!} \Gamma^{a aa}_{ 1 2 3 }   
a_{ 1} a_{ 2 } a_{3 } +  \frac{1}{3!} \Gamma^{\bar{a} \bar{a} \bar{a}}_{ 1 2 3 }   
a^{\dagger}_{ -1} a^{\dagger}_{ -2 } a^{\dagger}_{- 3 } 
 \Bigr],
 \label{eq:H3def}
 \end{eqnarray}
where $N$ is the total number of sites of the underlying lattice.
For simplicity we write $a_1 \equiv a_{ \bd{k}_1 } $ and 
abbreviate the interaction vertices by $\Gamma^{\bar{a}aa}_{ \bd{k}_1 ;
 \bd{k}_2 \bd{k}_3} \equiv \Gamma^{\bar{a}aa}_{ 1 ;
 2 3}$ etc..
Finally, the part $H_4$ to the Hamiltonian involving
four powers of the boson operators is in the absence of $U(1)$-symmetry given by
\begin{eqnarray}
 H_4 & = & \frac{1}{N} \sum_{ \bd{k}_1 \ldots \bd{k}_4} 
 \delta_{ \bd{k}_1 + \ldots + \bd{k}_4 ,0}
 \Bigr[ \frac{1}{(2!)^2} \Gamma^{ \bar{a} \bar{a} aa}_{ 12; 3 4}   
a^{\dagger}_{ -1} a^{\dagger}_{-2} a_{ 3 } a_{4 } 
 \nonumber
 \\
 &+ &
\frac{1}{3!} \Gamma^{ \bar{a} aaa}_{ 1; 2 3 4}   
a^{\dagger}_{ -1} a_{ 2 } a_{3 } a_4
 +  \frac{1}{3!} \Gamma^{ \bar{a} \bar{a} \bar{a}a}_{ 1 2 3; 4 }   
a^{\dagger}_{ -1} a^{\dagger}_{ -2 } a^{\dagger}_{-3} a_{4 } 
 \nonumber
 \\
 &+ &
  \frac{1}{4!} \Gamma^{a aaa}_{ 1 2 3 4}   
a_{ 1} a_{ 2 } a_{3 } a_4 +  \frac{1}{4!} \Gamma^{\bar{a} \bar{a} \bar{a} \bar{a}}_{ 1 2 3 4}   
a^{\dagger}_{ -1} a^{\dagger}_{ -2 } a^{\dagger}_{- 3 } a^{\dagger}_{-4}
 \Bigr]. \hspace{7mm}
 \label{eq:H4def}
 \end{eqnarray}
In the appendix we shall show how to obtain a boson Hamiltonian of the above form
from an effective spin Hamiltonian
describing the lowest magnon band
of YIG in the so-called parallel pumping geometry~\cite{Kreisel09,Tupitsyn08,Kloss10}.

The spatial dependence of the Bose condensate is determined
by the Gross-Pitaevskii equation~\cite{Pitaevskii03}, which can be obtained
from the extremum of the corresponding Euclidean action.
In order to write the various interaction processes 
in a compact notation, we introduce a two-component complex field 
$\Phi^{\sigma}_{ \bd{k}} ( \tau )$, where $\tau$ is the imaginary time and
$\sigma = a, \bar{a}$ labels the two components according to the prescription
 \begin{equation}
 a_{\bd{k}} \rightarrow \Phi^a_{ \bd{k}} ( \tau ) \; \; , \; \;
 a^{\dagger}_{\bd{k}} \rightarrow \Phi^{\bar{a}}_{ -\bd{k}} ( \tau ).
 \end{equation}
The quadratic part $H_2$ of the Hamiltonian corresponds then to the Gaussian 
action
 \begin{eqnarray}
 S_2 [ \Phi ] & = &   \frac{1}{2} \int_0^{\beta} d \tau   \sum_{\bd{k}}  
(   \Phi^{\bar{a}}_{- \bd{k}} , \Phi^a_{ - \bd{k}} )
 \nonumber
\\
 &\times & 
 \left( \begin{array}{cc}
 \partial_{\tau} + \epsilon_{\bd{k}}  - \mu & \gamma_{\bd{k}}
 \\
\gamma_{\bd{k}}^{\ast} & -\partial_{\tau} + \epsilon_{\bd{k}} - \mu
 \end{array} \right)
\left( \begin{array}{c}
\Phi^a_{\bd{k}} \\
 \Phi^{\bar{a}}_{\bd{k}}
 \end{array} \right), \hspace{7mm}
 \end{eqnarray}
where $\beta$ is the inverse temperature and $\mu$ is the chemical potential.
The Euclidean action corresponding to the
interaction parts $H_3$ and $H_4$ can be written in the following symmetrized form,
 \begin{eqnarray}
S_3 [ \Phi ] & = &
 \int_0^{\beta} d \tau
  \frac{1}{\sqrt{N}  }    \sum_{ \bd{k}_1 \bd{k}_2 \bd{k}_3}
\sum_{ \sigma_1 \sigma_2 \sigma_3 } 
 \delta_{ \bd{k}_1 + \bd{k}_2 + \bd{k}_3 , 0}
  \nonumber
 \\
 & & \times \frac{1}{3!}
\Gamma_3 ( \bd{k}_1 \sigma_1 , \bd{k}_2 \sigma_2 , \bd{k}_3 \sigma_3 )
 \Phi^{\sigma_1}_{\bd{k}_1} \Phi^{\sigma_2}_{\bd{k}_2} \Phi^{\sigma_3}_{\bd{k}_3},
 \\
S_4 [ \Phi ] & = &
 \int_0^{\beta} d \tau
   \frac{1}{N }    \sum_{ \bd{k}_1 \ldots \bd{k}_4}
\sum_{ \sigma_1 \ldots \sigma_4 } 
 \delta_{ \bd{k}_1 + \ldots + \bd{k}_4 , 0}
  \nonumber
 \\
 & & \times \frac{1}{4!}
\Gamma_4 ( \bd{k}_1 \sigma_1 , \ldots , \bd{k}_4 \sigma_4 )
 \Phi^{\sigma_1}_{\bd{k}_1} \ldots \Phi^{\sigma_4}_{\bd{k}_4},
\end{eqnarray} 
where the flavor indices $\sigma_i = a , \bar{a}$ keep track of the
two different field types, and
the vertices 
$\Gamma_3 ( \bd{k}_1 \sigma_1 , \bd{k}_2 \sigma_2 , \bd{k}_3 \sigma_3 )$ and
$\Gamma_4 ( \bd{k}_1 \sigma_1 , \ldots , \bd{k}_4 \sigma_4 )$
are completely symmetric under the permutation of all indices.
The combinatorial factors
in these expressions are chosen~\cite{Schuetz05} such that for a given 
ordering of the indices the completely symmetrized vertices
can be identified with the partially symmetrized vertices
appearing in Eqs.~(\ref{eq:H3def},~\ref{eq:H4def}), 
for example
 \begin{equation}
 \Gamma_3 ( \bd{k}_1 \bar{a} , \bd{k}_2 a , \bd{k}_3 a ) = \Gamma^{ \bar{a} aa}_{ 1; 23 }.
 \end{equation}
In the presence of a Bose condensate some of the expectation values
$\phi_{\bd{k}}^{\sigma} = \langle \Phi_{\bd{k}}^{\sigma} \rangle$ are finite and proportional to $\sqrt{N}$.
In equilibrium the order parameter fields $\phi_{\bd{k}}^{\sigma}$ are independent of the
imaginary time.
It is then useful to shift the integration variables $\Phi$ in the Euclidean functional
integral according to $\Phi_{\bd{k}}^{\sigma} ( \tau ) = \phi_{\bd{k}}^{\sigma} 
 + \delta \Phi_{\bd{k}}^{\sigma} ( \tau )$ and expand the
Euclidean action $S [ \Phi ] = S_2 [ \Phi ] + S_3 [ \Phi ] + S_4 [ \Phi ]$
in powers of the fluctuations,
 \begin{equation}
 S [ \phi + \delta \Phi ] = S [ \phi ] + \int_0^{\beta} d \tau \sum_{\bd{k} \sigma}
 \left. \frac{ \delta S [ \Phi ]}{\delta \Phi_{\bd{k}}^{\sigma} ( \tau ) } 
 \right|_{ \Phi=\phi } \delta  \Phi_{\bd{k}}^{\sigma} ( \tau ) + \ldots.
 \end{equation}
The physical order parameter field is determined by demanding that the
first variation of the action vanishes, which yields the Gross-Pitaevskii equation
 \begin{eqnarray}
 0 & = &  \left. \frac{ \delta S [ \Phi ]}{\delta \Phi_{\bd{k}}^{\sigma} ( \tau ) } 
 \right|_{ \Phi= \phi } 
  =  ( \epsilon_{\bd{k}} - \mu ) \phi^{\bar{\sigma}}_{-\bd{k}}  + \gamma_{\bd{k}}^{\sigma}
 \phi^{{\sigma}}_{-\bd{k}}
 \nonumber
 \\
 & + &  \frac{1}{\sqrt{N}  }    \sum_{ \bd{k}_1 \bd{k}_2}
\sum_{ \sigma_1 \sigma_2  } 
 \delta_{ \bd{k} + \bd{k}_1 + \bd{k}_2 , 0}
  \nonumber
 \\
 & &  \times \frac{1}{2}
\Gamma_3 ( \bd{k} \sigma , \bd{k}_1 \sigma_1 , \bd{k}_2 \sigma_2 )
 \phi^{\sigma_1}_{\bd{k}_1} \phi^{\sigma_2}_{\bd{k}_2}
 \nonumber
 \\
&+  & 
   \frac{1}{N }    \sum_{ \bd{k}_1 \bd{k}_2 \bd{k}_3}
\sum_{ \sigma_1 \sigma_2 \sigma_3 } 
 \delta_{ \bd{k} + \bd{k}_1 + \bd{k}_2 + \bd{k}_3 , 0}
  \nonumber
 \\
 & &  \times \frac{1}{3!}
\Gamma_4 ( \bd{k} \sigma , \bd{k}_1 \sigma_1 , \bd{k}_2 \sigma_2,  \bd{k}_3 \sigma_3 )
 \phi^{\sigma_1}_{\bd{k}_1}  \phi_{\bd{k}_2}^{\sigma_2 } \phi^{\sigma_3}_{\bd{k}_3},
 \hspace{7mm}
 \label{eq:GP}
\end{eqnarray} 
where we have defined $\gamma_{\bd{k}}^{a} = \gamma_{\bd{k}}^{\ast}$ and
$\gamma_{\bd{k}}^{\bar{a}} = \gamma_{\bd{k}}$.

\begin{figure}[tb]
  \centering
\includegraphics[width=70mm]{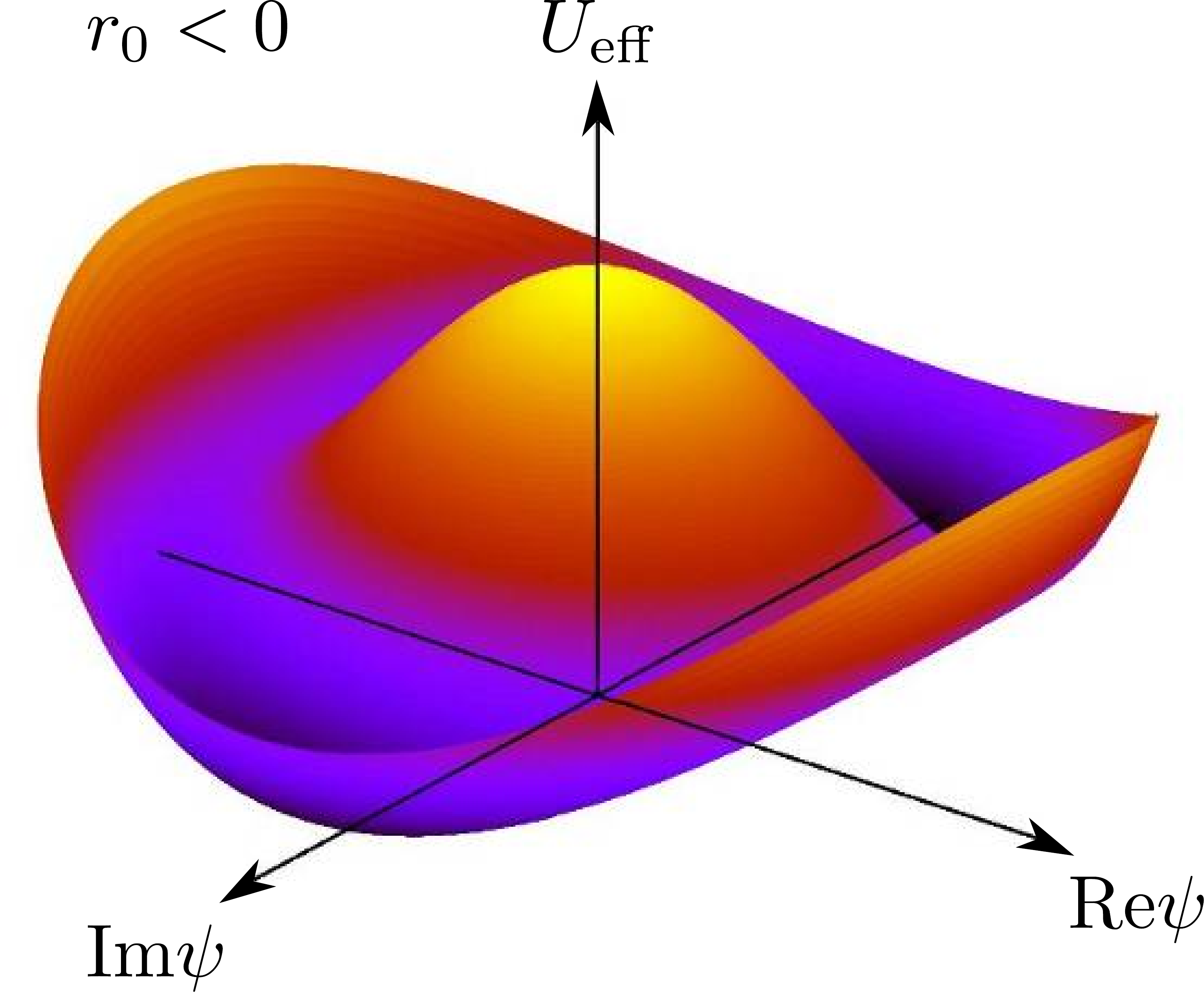}
\vspace{10mm}

\includegraphics[width=70mm]{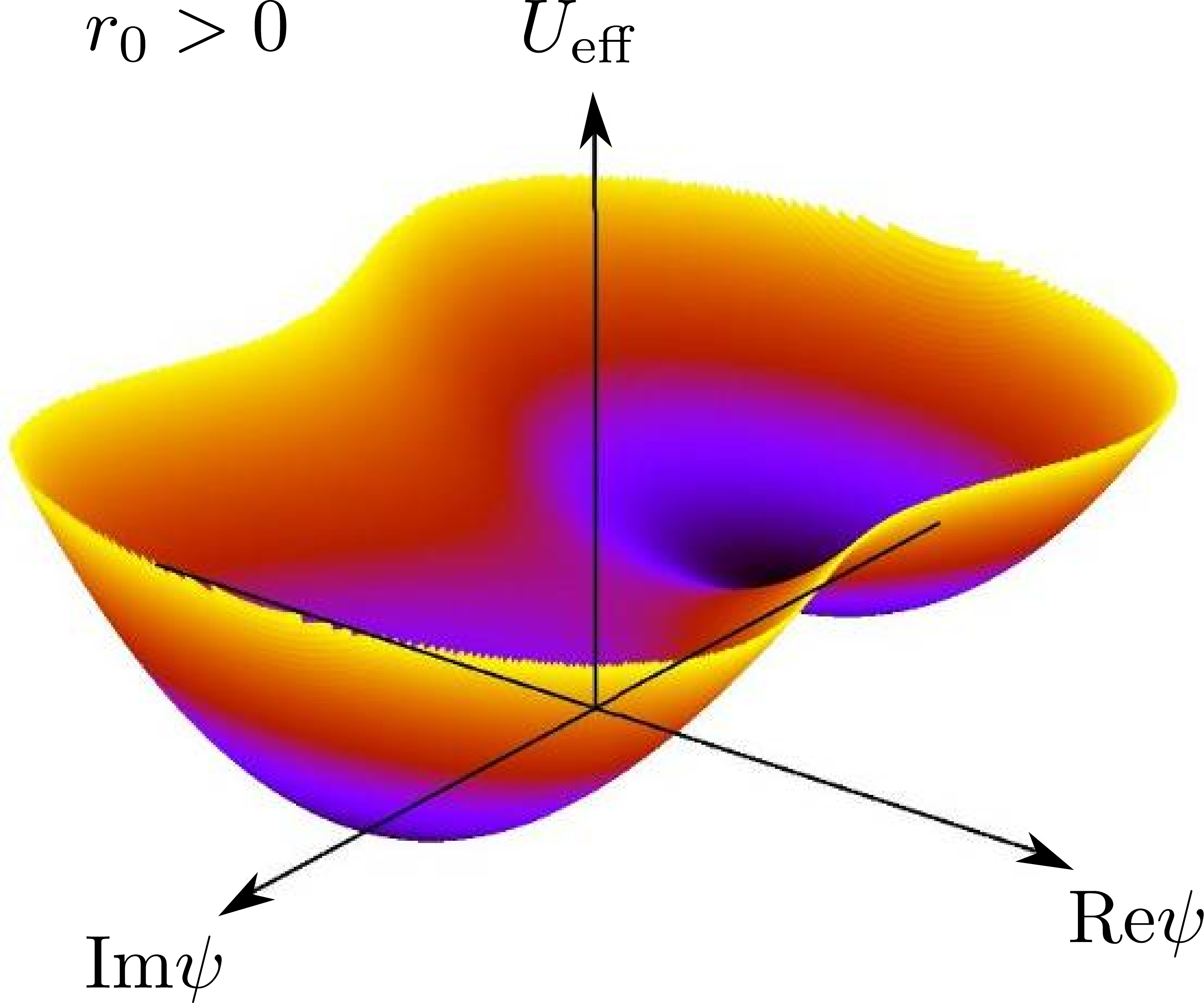}
  \caption{
(Color online)
Effective potential
for zero momentum BEC in the presence of
explicit symmetry breaking, see Eq.~(\ref{eq:Ueff}).
 The quadratic part of the Hamiltonian
is then given by Eq.~(\ref{eq:H2def}), where for simplicity we assume that
$\gamma_0$ is real and positive.
For the graphs the cubic vertices have been neglected and only the particle-number
conserving component of the four-point vertices
$\Gamma^{\bar{a} \bar{a} aa}_{ 00;00} \equiv u_4$
has been retained.
The graphs are for $\gamma_0 - r_0 > 0$ where the effective potential
has two degenerate minima on the imaginary axis.
Upper graph: for $ r_0 < 0$ the center of the effective potential is a local maximum 
so that its shape resembles Napoleon's hat.
Lower graph: for $ r_0 > 0$  the local maximum in (a) transforms into a saddle point.
}
  \label{fig:Napoleon}
\end{figure}

To begin with, let us assume that the system condenses in a state
where only the $\bd{k}=0$ mode is macroscopically occupied.
Such a state tends to be favored if
the dispersion has a minimum at $\bd{k} =0$.
In this case
 \begin{equation}  
 \phi^{\sigma}_{\bd{k}}  = \delta_{\bd{k} , 0} \sqrt{N} \psi^{\sigma}_0,
 \end{equation}
 where
the complex parameter $\psi^{a}_0 = ( \psi^{\bar{a}}_0 )^{\ast}$
is  expected to be of the order of unity.
Assuming for simplicity that $\gamma_0^{\sigma} = \gamma_0$ is real, we then obtain
from our general Gross-Pitaevskii equation (\ref{eq:GP})
 \begin{eqnarray}
 0 &= &
   r_0 \psi^{\bar{\sigma}}_{0}  + \gamma_{0}
 \psi^{{\sigma}}_{0} +
  \frac{1}{2} \sum_{ \sigma_1 \sigma_2}
\Gamma_3 ( 0 \sigma , 0 \sigma_1 , 0 \sigma_2 )
 \psi^{\sigma_1}_{0} \psi^{\sigma_2}_{0}
 \nonumber
 \\
&+ &  
     \frac{1}{3!} \sum_{ \sigma_1 \sigma_2 \sigma_3}
\Gamma_4 ( 0 \sigma , 0 \sigma_1 , 0 \sigma_2,  0 \sigma_3 )
 \psi^{\sigma_1}_{0}  \psi_{0}^{\sigma_2 } \psi^{\sigma_3}_{0},
 \hspace{7mm}
 \label{eq:GP0}
\end{eqnarray} 
where 
 \begin{equation}
r_0 = \epsilon_0 - \mu.
 \end{equation}
We adopt here the standard notation
in the field of critical phenomena~\cite{Ma76} where a negative value of $r_0$
implies a finite expectation value of the order parameter field.
Assuming further that
the three-legged vertices vanish and that
only the particle number conserving four-legged vertex
$\Gamma^{\bar{a} \bar{a} aa}_{ 00;00} \equiv u_4$ is finite and positive,
we find that the Gross-Pitaevskii equation (\ref{eq:GP0}) has two
degenerate solutions
 \begin{equation}
 \psi_0^{\sigma} = \pm i \sqrt{ \frac{2 (  \gamma_0 - r_0)  }{u_4}}.
 \label{eq:phi0res} 
\end{equation}
Note that even for positive $r_0 = \epsilon_0 - \mu$
the condensate is stable if the energy scale $\gamma_0$ associated
with explicit symmetry breaking  is sufficiently large.
Of course, for $\gamma_0 \neq 0$ there is no spontaneous symmetry breaking so that
there is no gapless Goldstone mode in the condensed state.
Due to the symmetry breaking terms $a^{\dagger}_{ \bd{k}} a^{\dagger}_{-\bd{k}}$
and $a_{-\bd{k}} a_{\bd{k}}$ in the quadratic part $H_2$ of our Hamiltonian
the effective potential 
 \begin{equation}
 U_{\rm eff} [ \psi_0^{\bar{a} } , \psi_0^{a} ] = N^{-1} 
S[ \Phi_{\bd{k}}^{\sigma} \rightarrow \sqrt{N} \delta_{ \bd{k},0} \psi_0^{\sigma} ]
 \label{eq:Ueff}
 \end{equation}
has two degenerate minima at purely imaginary
values of the field as shown 
in Fig.~\ref{fig:Napoleon}. 
Cubic terms (which are neglected in these plots)
distort the effective potential and break the degeneracy of the two minima.
The problem of BEC at zero momentum
in the presence of $U(1)$ symmetry breaking terms has been
discussed previously in Ref.~\cite{DellAmore09}.

\begin{figure}[t]
\includegraphics[width=80mm]{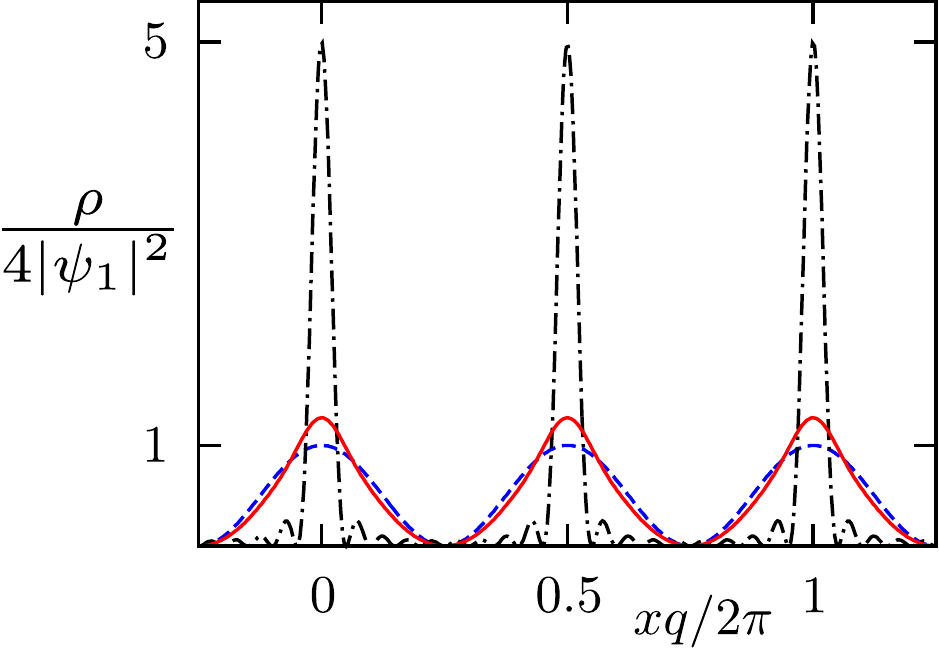}
  \caption{
(Color online)
This plot illustrates the fact that the condensate wave-function is more localized in real space
if many Fourier components contribute.
The dash-doted and the solid lines represent condensate
densities associated with a condensate wave-function
where the $5$ odd Fourier modes $\bd{q}_1, \bd{q}_3, \bd{q}_5, \bd{q}_7, \bd{q}_9$ are
macroscopically occupied. For the dashed-dotted line we have assumed that all 5 modes have the same 
weight, while the dashed line corresponds to YIG with pumping parameter $\gamma_1/r_1=3$, see Table \ref{tab:YIG}.
For comparison, the dashed line is the condensate density $\rho_1\left(\bd{r}\right)$ where only the 
modes with wave vectors $\pm\bd{q}$ are macroscopically occupied, see Eq. \ref{eq:dens1}.
}
  \label{fig:cdensity}
\end{figure}

Next let us study the more interesting case where
the dispersion $\epsilon_{\bd{k}}$ has two degenerate minima 
at finite wave-vectors $\pm {\bd{q}}$. At the first sight it seems that
in this case one can find
solutions of the  Gross-Pitaevskii equation (\ref{eq:GP}) 
where only the modes with $\bd{k} = \pm \bd{q}$ condense,
 \begin{equation}
 \phi^{\sigma}_{\bd{k}} = \sqrt{N} [ 
 \delta_{ \bd{k} , \bd{q}} \psi^{\sigma}_1 + \delta_{ \bd{k} , - \bd{q}} \psi^{\sigma}_{-1}].
 \label{eq:condensq}
 \end{equation}
Keeping in mind that
 $\psi_1^{a} = \langle a_{\bd{q}} \rangle / \sqrt{N}$ 
and $\psi_{-1}^{\bar{a}} = \langle a^{\dagger}_{\bd{q}} \rangle/ \sqrt{N}$,
we see that  
 $(\psi^{{\sigma}}_{1} )^{\ast} = \psi^{\bar{\sigma}}_{-1}$.
In real space  the condensate wave-function (\ref{eq:condensq}) corresponds to
 \begin{equation}
 \phi^{\sigma} ( \bd{r} ) = \frac{1}{\sqrt{N}} \sum_{\bd{k}} 
 e^{ i \bd{k} \cdot \bd{r}} \phi_{\bd{k}}^{\sigma} =  e^{ i {\bd{q}} \cdot 
 {\bd{r}} } \psi_1^{\sigma}  +   e^{ - i {\bd{q}} \cdot 
 {\bd{r}} } \psi_{-1}^{\sigma}.
 \label{eq:sol1}
 \end{equation}
Setting $\psi_1^{\sigma} = 
\psi_{-1}^{\sigma} = \psi $, the corresponding
condensate density is
 \begin{equation}
  \rho_1 ( \bd{r} ) = | \phi^{{a}} ( \bd{r} ) |^2 =
4 | \psi |^2 \cos^2 ( \bd{q} \cdot \bd{r} ).
 \label{eq:dens1}
 \end{equation}
The important point is now that  a condensate wave-function of this type 
does {\it{not}} solve the Gross-Pitaevskii equation (\ref{eq:GP}), because
the interaction terms couple the Fourier components with $ \bd{k} = \pm \bd{q}$
to all other Fourier components involving  arbitrary integer multiples $ n \bd{q}$
of the fundamental wave-vector $\bd{q}$, where $n=0, \pm 1, \pm 2 , \ldots$.
To see this more clearly, let us substitute the general ansatz
 \begin{equation}
 \phi^{\sigma}_{\bd{k}} = \sqrt{N}   \sum_{n=-\infty}^{\infty} \delta_{ \bd{k} ,  \bd{q}_n}
\psi^{\sigma}_n 
 \label{eq:congen}
 \end{equation}
into Eq.~(\ref{eq:GP}) where $\bd{q}_n = n \bd{q}$.
Setting the external wave-vector $ \bd{k} = - \bd{q}_n$  in 
Eq.~(\ref{eq:GP}) and defining $r_n = \epsilon_{ - \bd{q}_n } - \mu$ and
$\gamma_n = \gamma_{ - \bd{q}_n }$ (assuming again that $\gamma_{\bd{k}}$ is real),
we obtain the discrete Gross-Pitaevskii equation,
 \begin{eqnarray}
 - r_n \psi_n^{ \bar{\sigma}} - \gamma_n \psi_n^{\sigma} & = & \frac{1}{2}
 \sum_{ n_1 n_2} \sum_{\sigma_1   \sigma_2 }  \delta_{ n, n_1 + n_2 } 
 V_{ n  n_1 n_2 }^{\sigma   \sigma_1 \sigma_2} \; 
 \psi^{\sigma_1}_{n_1} \psi^{\sigma_2}_{n_2}  
 \nonumber
 \\
 & & \hspace{-30mm}  + \frac{1}{3!} \sum_{ n_1 n_2 n_3} \sum_{\sigma_1   \sigma_2 \sigma_3}  \delta_{ n, n_1 + n_2 +n_3}  
U_{ n  n_1 n_2 n_3}^{\sigma   \sigma_1 \sigma_2 \sigma_3} \;
 \psi^{\sigma_1}_{n_1} \psi^{\sigma_2}_{n_2}  \psi^{\sigma_3}_{n_3}  ,
 \label{eq:GPdis}
 \end{eqnarray}
where
 \begin{eqnarray}
 V_{ n  n_1 n_2 }^{\sigma   \sigma_1 \sigma_2} & = & \Gamma_3 ( \bd{q}_n \sigma, 
 \bd{q}_{n_1} \sigma_1 , \bd{q}_{n_2} \sigma_2 ),
 \label{eq:V3}
 \\
U_{ n  n_1 n_2 n_3}^{\sigma   \sigma_1 \sigma_2 \sigma_3} & = & 
\Gamma_4 ( \bd{q}_n \sigma, 
 \bd{q}_{n_1} \sigma_1 , \bd{q}_{n_2} \sigma_2, \bd{q}_{n_3} \sigma_3 ).
 \label{eq:U4}
 \end{eqnarray}
The crucial point is now that if we
assume on the right-hand side
of Eq.~(\ref{eq:GPdis}) that only the coefficients $\psi^{\sigma_i}_{ n_i}$ with $n_i = \pm 1$
are finite, then we find after carrying out the sum
that on the left-hand side all 
field components $\psi_n^{\sigma}$ with
$n=0, \pm 1, \pm 2, \pm 3$ must also be finite, so that
the assumption that only the modes with wave-vector $\pm \bd{q}$
condense is not self-consistent. 
For general interactions where all
interaction coefficients
 $V_{ n  n_1 n_2 }^{\sigma   \sigma_1 \sigma_2}$
and
$U_{ n  n_1 n_2 n_3}^{\sigma   \sigma_1 \sigma_2 \sigma_3}$ are finite,
the Fourier transform of
a self-consistent solution of the Gross-Pitaevskii equation
must therefore have finite weight for all integer multiples of
$\bd{q}$.
Depending on the  behavior of the interaction coefficients
the spatial behavior of the condensate wave-function can look rather differently. 
In Sec.~\ref{sec:YIG} we shall show that for YIG the cubic interaction coefficients
$V_{ n  n_1 n_2 }^{\sigma   \sigma_1 \sigma_2}$ actually vanish identically,
and that the behavior of the quartic coefficients
$U_{ n  n_1 n_2 n_3}^{\sigma   \sigma_1 \sigma_2 \sigma_3}$ is such that
the component $\psi_{\pm 1}^{\sigma}$ of the condensate wave-function
is much larger than the other components.
In this case the spatial distribution of the condensate density
is to a good approximation given by  Eq.~(\ref{eq:dens1}).
On the other hand for some other types of interactions many Fourier components
of the solution of the discrete Gross-Pitaevskii equation can have the same 
order of magnitude. In this case the condensate density
is strongly localized at the sites of a one-dimensional lattice with  spacing
$2 \pi / | \bd{q} |$. For example, if we assume that 
the first $m$ Fourier components are finite and equal,
$\psi^{\sigma}_n = \psi / \sqrt{m}$ for $ 1 \leq | n | \leq m$, then the condensate density
is given by
 \begin{equation}
 \rho_m ( \bd{r} ) =  \left|
\frac{ 2  \psi}{\sqrt{m}}  \sum_{ n =1}^m \cos ( \bd{q}_n \cdot \bd{r} ) 
 \right|^2.
 \label{eq:rhoM}
 \end{equation}
To obtain a properly normalized density, it is necessary to scale the
order parameter as $ \psi / \sqrt{m}$.
In Fig.~\ref{fig:cdensity} we compare 
the single-component density $\rho_1 ( \bd{r} )$ given in
Eq.~(\ref{eq:dens1}) with the corresponding 
multi-component density where the $5$ odd Fourier modes $\bd{q}_1, \bd{q}_3, \bd{q}_5, \bd{q}_7, \bd{q}_9$ are
macroscopically occupied.
Obviously, in this case one can already observe a strong localization
of the condensate at the sites of a one-dimensional lattice with spacing
$\pi / | \bd{q} |$.  It is then appropriate to think of 
BEC as a condensation phenomenon in real space.
In fact, the formation of the Bose condensate resembles in this case the 
phase transition from a liquid to a crystalline 
solid~\cite{Yukalov78}.
However, in the three-dimensional crystal formation problem the 
situation is more complicated because 
the Gaussian term in a Ginzburg-Landau theory 
exhibits a minimum on a  surface in momentum space, and for the crystal structure
the cubic term in the expansion of the Landau functional
in powers of the density play also an important role~\cite{Alexander78,Anderson84}.

\section{BEC of magnons in YIG}
\label{sec:YIG}

\begin{figure}[tb]
\centering{}
\includegraphics[scale=0.8]{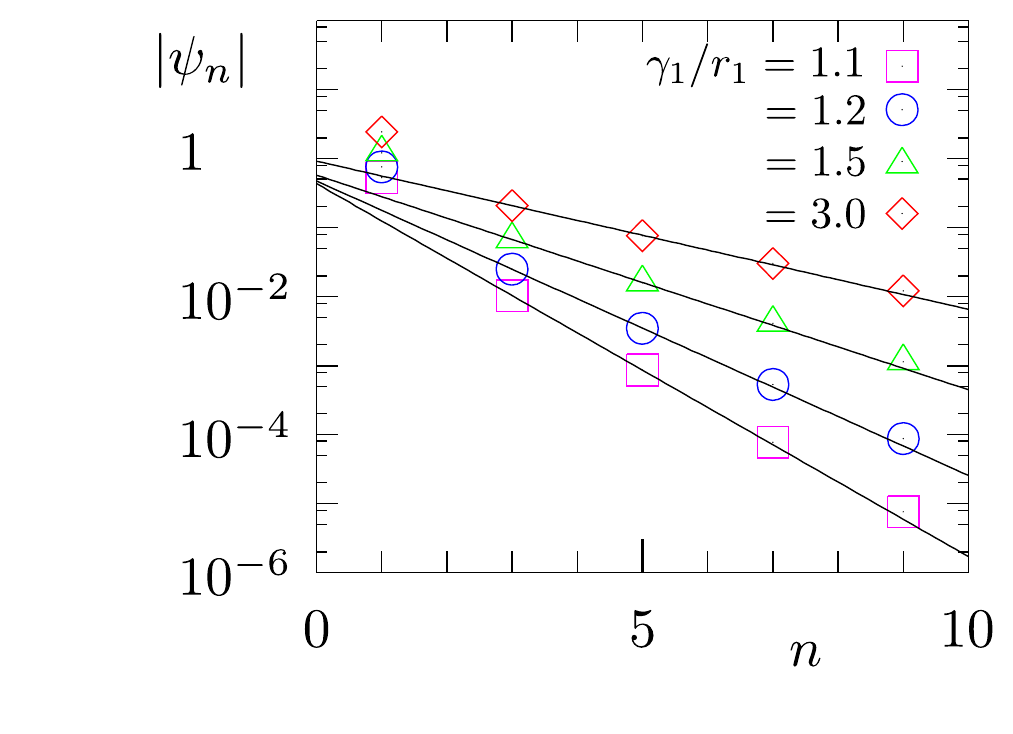}
\caption{ (Color online) Absolute values
 $| \psi_n | = \left|\psi_{n}^{\sigma} \right|$
of the Fourier components  of the order parameter 
for BEC in YIG for different 
values of the dimensionless ratio  $\gamma_1 / r_1 = 
\gamma_{\bd{q}}/( \epsilon_{\bd{q}} -  \mu)$.
The order parameter has been obtained from the numerical solution
of the discrete Gross-Pitaevskii equation  (\ref{eq:GPdis}), using the fact that
for YIG the three-point vertices $V^{\sigma \sigma_1 \sigma_2}_{ n n_1 n_2}$ vanish.
In this case the physically relevant solution of Eq.~(\ref{eq:GPdis}) has only odd Fourier
components. Note the Fourier components $\psi_1^{\sigma}$ are dominant and
that the higher order Fourier components $\psi_n^{\sigma}$
decay approximately exponentially as a function of $n$.
\label{fig:order}}
\end{figure}

It is well known~\cite{Kreisel09,Kloss10,Rezende09,Lvov94,Zakharov70,Tsukernik75,Vinikovetskii79,Lim88,Kalafati89}
 that the effective magnon Hamiltonian for
YIG can be cast into the general form given in Eqs.~(\ref{eq:Hdef}--\ref{eq:H4def}).
In the appendix we summarize the main steps and approximations
in the derivation of the magnon Hamiltonian for YIG 
from a realistic spin Hamiltonian and give
explicit expressions for the interaction vertices.
If the samples have the shape of thin stripes and if an external magnetic field
is oriented along the direction of the stripes (which we call the $z$-direction), the
energy dispersion $\epsilon_{\bd{k}}$ of the lowest magnon band 
indeed has two degenerate minima wave-vectors $\pm \bd{q}=
 \pm q \bd{e}_z$. From the explicit expressions for the three-point vertices
for YIG
given in Eqs. (\ref{eq:Gammabbb1}, \ref{eq:Gammabbb2}) and (\ref{eq:Gammaaaa1}--\ref{eq:Gammaaaa4}) we see that for this direction of $\bd{q}$
all three-magnon interaction vertices $V^{\sigma \sigma_1 \sigma_2}_{ n n_1 n_2 }$
defined in Eq.~(\ref{eq:V3}) vanish identically, so that
for the discussion of BEC in YIG one can omit the
first term on the right-hand side of the discrete Gross-Pitaevskii
equation (\ref{eq:GPdis}).
We can then construct self-consistent solutions of this equation
involving only  Fourier components $\psi_{\pm n}^{\sigma}$ with $n = 
( 2 j+1) n_1$, where $j =0,1,2, \ldots$, and $n_1 $ is the label of the lowest finite
Fourier component.
Because in experiments the samples are kept at room temperature, leading to a finite thermal magnon
density at ${\bd k}=\pm \bd{q}$, and energy transfer by pumping is mostly done to modes whose energy is less than $\epsilon_{2{\bd q}}$, we expect that the Fourier components
$ \psi^{\sigma}_{\pm 1}$ are dominant. In the following we therefore set $n_1 =1$ and consider solutions
of the type
 \begin{equation}
 \phi^{\sigma}_{\bd{k}} = \sqrt{N} \sum_{ n \; \rm{ odd}} \delta_{\bd{k} , \bd{q}_{n} }
 \psi^{\sigma}_{n}\;.
 \label{eq:psiodd}
 \end{equation}
The infinite set of Fourier components $\psi_{\pm 1}^{\sigma}, \psi_{\pm 3}^{\sigma},
 \psi_{\pm 5}^{\sigma} , \ldots$
is determined by setting $n = \pm 1, \pm 3 , \pm 5, \ldots$
in the discrete Gross-Pitaevskii equation (\ref{eq:GPdis}), keeping in mind that
for BEC of magnons in YIG we should set 
 $V^{\sigma \sigma_1 \sigma_2}_{ n n_1 n_2 }~=~0$ and use the
four-point vertices
$U_{ n  n_1 n_2 n_3}^{\sigma   \sigma_1 \sigma_2 \sigma_3}$ defined
via Eqs.~(\ref{eq:U4}), (\ref{eq:UU1}--\ref{eq:UU3}), and (\ref{eq:Uaa1}--\ref{eq:Uaa5}).
We have solved these equations numerically by truncating the
expansion (\ref{eq:psiodd}) at some finite order $m> | n |$. 
For positive $r_1$ non-trivial solutions can be obtained for
 \begin{equation}
 \gamma_1 > r_1 = \epsilon_{\bd{q}} - \mu\;.
 \end{equation}
If $r_1$ is negative, we find solutions for arbitrary $\gamma_1$, including $\gamma_1 =0$.
As discussed in the appendix, to describe the stationary non-equilibrium state
of the magnon gas in YIG under the influence of an external microwave field
oscillating with frequency $\omega_0$, one should re-define
$\epsilon_{\bd{q}} \rightarrow \epsilon_{\bd{q}} - \omega_0/2$ and
use an appropriate chemical potential $\mu$.
It turns out that the Fourier coefficients $ \psi_n^{\sigma}$
decay rapidly for large $n$, so that in practice it is not necessary to choose 
the cutoff $m$ larger than $10$ to obtain converged results.
Typical numerical results are summarized in Table I and are represented graphically
in Fig.~\ref{fig:order}.
\begin{table}

\centering{}\caption{\label{tab:YIG} Numerical results for $\left|\psi_{1}\right| \equiv | \psi_1^{\sigma} |$
and the ratios $\left|\psi_{n}\right|/ \left|\psi_{1}\right|$
for different values of the dimensionless pumping parameters $\gamma_1 / r_1$ for $r_1>0$. Note that for $r_1>0$ there is no condensate if $\gamma_1 / r_1<1$.
The numbers have been obtained from the numerical solution of the
discrete Gross-Pitaevskii equation (\ref{eq:GPdis}) using realistic interaction
parameters for YIG. By symmetry, for negative momenta we obtain identical results.}
\begin{tabular}{cccccc}
$\gamma_{1}/r_{1}$ & $\left|\psi_{1}\right|$ & $\left|\psi_{3}/\psi_{1} \right|$ & $\left|\psi_{5}/\psi_{1} \right|$ & $\left|\psi_{7}/\psi_{1} \right|$ & $\left|\psi_{9}/\psi_{1} \right|$\tabularnewline
\hline 
$1.1$ & $0.530$ & $0.019$ & $0.002$ & $0$ & $0$\tabularnewline
$1.2$ & $0.757$ & $0.033$ & $0.005$ & $0.001$ & $0$\tabularnewline
$1.5$ & $1.209$ & $0.055$ & $0.013$ & $0.003$ & $0.001$\tabularnewline
$3.0$ & $2.439$ & $0.085$ & $0.031$ & $0.012$ & $0.005$\tabularnewline
\end{tabular}
\end{table}

Obviously, for YIG the first Fourier components $\psi^{\sigma}_{ \pm 1}$
of the condensate wave function are dominant, so that
the spatial structure of the condensate density is
to a good approximation given by $\rho_1 ( \bd{r} )$
given in Eq.~(\ref{eq:dens1}).
This is consistent with the experiments by 
Demokritov and co-workers~\cite{Demokritov06,Demidov07,Dzyapko07,Demidov08,Demokritov08}, 
who observed a strong enhancement of the magnon distribution only
for wave-vectors $\pm \bd{q}$. Note, however, that
in principle also the higher order Fourier components are finite.
In fact, 
with increasing amplitude of the pumping field
the relative weight of the higher order Fourier components also increases.
For example, from our numerical data shown in Table~I we see that for
$ \gamma_1 / r_1 = 3$ the amplitude of the 
third Fourier component $\psi_{\pm 3}^{\sigma}$
is approximately $8 \%$ of the amplitude of the dominant component
$\psi_{\pm 1}^{\sigma}$. 
If the first Fourier components have the same order of magnitude and one uses realistic
parameters for YIG, we estimate that the effective lattice spacing is 
$2\pi/|{\bd q}|\approx 2\cdot 10^{-7}\;\text{m}$, which is roughly a factor of 100
larger than the spacing of the underlying Bravais lattice.

\section{Summary and conclusions}

In summary, we have considered the general problem of BEC in an interacting
Bose gas whose energy dispersion has two degenerate minima at finite wave-vectors
$\pm \bd{q}$. 
Our main result is that for generic interactions the
condensate wave-function has finite Fourier components $\psi_n$ for all 
integer multiples $\bd{q}_n = n \bd{q}$ of the fundamental wave-vector $\bd{q}$.
For special interactions many Fourier components $\psi_n$ can have the same
order of magnitude, so that in real space the condensate is strongly localized
at the sites of a one-dimensional lattice. In this case there is a formal analogy
between BEC and the liquid-solid transition.

We have also used our theory to study  the condensate wave-function
of the condensed magnons in the magnetic insulator yttrium-iron garnet.
In this case it is appropriate to think of magnon BEC as a condensation process
in momentum space, because the condensate wave-function is
dominated by its leading Fourier components at $\pm {\bd{q}}$.
However, if the amplitude of the  oscillating external microwave
field is increased, then higher order Fourier components of the condensate wave-function can be 
populated.

If one succeeds to extend the experiments which observe directly the magnon densities 
to wavevectors $k \gtrsim q$ it should be possible to detect also these higher order 
components. Since the order of magnitude of the higher order Fourier components drops
exponentially and delocalization problems occur due to a finite slope of the dispersion 
away from the minimum, it is a rather challenging task to verify our prediction 
experimentally.

Finally, it should be mentioned that quite recently
Malomed {\it{et al.}} \cite{Malomed10}
studied the dynamics of BEC of magnons in YIG
within an approach where coupled time-dependent Gross-Pitaevskii type of equations
for the two components of the condensate wave-function corresponding to
condensation at $\pm \bd{q}$ are written down phenomenologically.
Note, however, that this approach neglects processes which couple the dominant 
Fourier components $\psi_{ \pm 1}$
to the higher order Fourier components of the condensate wave-function.

\section*{ACKNOWLEDGMENTS}
We thank A. Serga, S. Demokritov, and A. Slavin for discussions. 
Financial support by
SFB/TRR49 and the DAAD/CAPES PROBRAL program is gratefully
acknowledged.

\begin{appendix}
\renewcommand{\theequation}{A\arabic{equation}}
\section*{APPENDIX: MAGNON-MAGNON INTERACTIONS IN YIG}
\setcounter{equation}{0}

In this appendix we outline the derivation of an effective boson
Hamiltonian of the form given in Eqs.~(\ref{eq:Hdef}--\ref{eq:H4def})
describing the lowest magnon band of YIG, starting from
the following time-dependent spin 
Hamiltonian, \cite{Kreisel09,Tupitsyn08,Kloss10,Cherepanov93,Rezende06,Rezende09}
 \begin{eqnarray} 
{H}_{\rm YIG} ( t )  &=& -\frac{1}{2} \sum_{ij} \sum_{\alpha \beta} \big[ J_{ij} \delta^{\alpha \beta} + D_{ij}^{\alpha \beta}   \big] S_i^\alpha S_j^\beta \nonumber \\
            &&   - \big[ h_0 + h_1 \cos (\omega_0 t) \big]  \sum_i S_i^z,
 \label{eq:Hspin}
\end{eqnarray} 
where $\alpha, \beta = x,y,z$ label the three components of the
spin operators $S^{\alpha}_i$, and $i, j =1, \ldots , N$ enumerate the $N$ sites
$\bd{r}_i$ of a cubic lattice with lattice spacing $a~\approx~12.376$~\AA.
The exchange couplings  $J_{i j}=J({\bd r}_i - {\bd r}_j)$  have the value $J \approx 1.29$K
if ${\bd r}_i - {\bd r}_j$
connect nearest neighbor sites and vanish otherwise.
The Zeeman energy associated with a static
external magnetic field $H_e$ 
is denoted by $h_0=\mu H_e$, where $\mu = g \mu_B$ with the Bohr magneton
given by $\mu_B$. Setting $g=2$ we should work with an effective spin 
$S\approx 14.2$ as discussed in Refs.~\cite{Kreisel09,Tupitsyn08}.
The time-dependent part of Eq.~(\ref{eq:Hspin}) represents
the Zeeman energy induced by an external microwave field
oscillating with frequency $\omega_0$.
The energy scale  $h_1$ associated with the oscillating component of the
magnetic field is assumed to be small compared with $h_0$, so that both the static and the oscillating magnetic field point into the direction of the
macroscopic magnetization, which we call the $z$-direction (parallel pumping).
Finally, the matrix elements of the
dipolar tensor $D_{ij}^{\alpha \beta} = D^{\alpha \beta}(\bd{r}_i - \bd{r}_j ) $ are
 \begin{equation}
 D_{ij}^{\alpha \beta}  = (1- \delta_{ij})\frac{\mu^2}{|\bd{r}_{ij}|^3} \left[ 3\hat r_{ij}^\alpha \hat r_{ij}^\beta - \delta^{\alpha \beta}  \right],
\end{equation}  
where ${\bd r}_{ij}={\bd r}_i - {\bd r}_j$ and $\hat {\bd r}_{ij} = {\bd r}_{ij}/|{\bd r}_{ij}|$.

Because the experimentally relevant YIG stripes are several thousand lattice spacings
thick,  we may assume that for magnetic fields oriented along the direction of the
stripes the classical ground state is a saturated ferromagnet with
all spins pointing in the direction of the external magnetic field.
The components of the spin operators can then be expressed in terms of canonical boson
operators $b_i$ and $b_i^{\dagger}$ as follows~\cite{Holstein40},
\begin{subequations}
\begin{eqnarray}
S_i^+ &=&  \sqrt{2S} \sqrt{1  - \frac{b^{\dagger}_i b_i}{2S} } \; b_i 
=  \sqrt{2S} \left[  b_i - \frac{b_i^\dagger b_i b_i}{4S} + \ldots \right],
 \nonumber
 \\
  & &
\\
S_i^- &= & 
 \sqrt{2S} 	b_i^{\dagger} \sqrt{1  - \frac{b^{\dagger}_i b_i}{2S} } 
= \sqrt{2S} \left[  b_i^\dagger - \frac{b_i^\dagger b_i^\dagger b_i}{4S} + \ldots \right],  
 \nonumber
 \\
 & &
\\
S_i^z &=& S-b_i^\dagger b_i, 
\end{eqnarray}
\end{subequations}
where $S_i^+ = S_i^x + i S_i^y$ and  $S_i^- = S_i^x - i S_i^y$. 
Retaining terms up to fourth order in the boson operators we obtain
 \begin{equation}
 H_{\rm YIG} (t) = H_0 ( t ) + H_2 ( t ) + H_3 + H_4 + {\cal{O}} ( S^{-1/2} ),
 \end{equation}
where the boson-independent term and the term quadratic in the bosons 
are~\cite{Kreisel09,Kloss10}
 \begin{eqnarray}
 H_0 ( t ) & = & - \frac{S^2 }{2} \sum_{ij} \left[ J_{ij} + D^{zz}_{ij} \right] 
 \nonumber
 \\
 & &
 - NS [h_0+h_1 \cos(\omega_0 t)], 
 \label{eq:H0def} 
 \\
 {H}_2 ( t ) &=& \sum_{ij} 
 \left[ A_{ij} b_i^\dagger b_j + \frac{B_{ij}}{2} \left( b_i b_j + b_i^\dagger b_j^\dagger  \right)  \right] 
 \nonumber
 \\
 & &
+h_1 \cos (\omega_0 t)\sum_i b_i^\dagger b_i, 
 \end{eqnarray}
with coefficients given by
 \begin{eqnarray}
 A_{ij} &=& \delta_{ij} h_e + S(\delta_{ij} \sum_n J_{in} - J_{ij}  ) \nonumber \\
 &&+ S \left[ \delta_{ij} \sum_{n} D_{in}^{zz} - \frac{D_{ij}^{xx} + D_{ij}^{yy} }{2}  \right], \\
 B_{ij} &=& - \frac{S}{2} [D_{ij}^{xx} - 2iD_{ij}^{xy} - D_{ij}^{yy} ].
 \end{eqnarray} 
The cubic contribution $H_3$ to the boson Hamiltonian is of order $\sqrt{S}$ and
involves only the dipolar tensor,
 \begin{eqnarray}
 {H}_3 & =& \sqrt{\frac{S}{2}} \sum_{ij} \Bigl[ 
\left( D_{ij}^{zy}+iD_{ij}^{zx} \right) 
\Bigl( b_i^\dagger  b_j^\dagger b_j +
\frac{1}{4} b_i^\dagger b_i^\dagger b_i    \Bigr) 
+{\rm{h.c.}} \Bigr].\nonumber \\
 \end{eqnarray}
The quartic part of the boson Hamiltonian can be written as
\begin{eqnarray}
{H}_4 &=& -\frac{1}{2} \sum_{ij} J_{ij} \Bigl[ n_i n_j 
-\frac{1}{2}\left(b_i^\dagger b_j^\dagger b_j b_j  + \textrm{h.c.} \right) \Bigr] 
 \nonumber 
\\
 &+  &
 \frac{1}{2} \sum_{ij} \left(D_{ij}^{xx}+D_{ij}^{yy}\right) \Bigl[ n_i n_j + \frac{1}{4}\left( b_i^\dagger b_j^\dagger b_j b_j + \textrm{h.c.} \right) \Bigr] 
\nonumber 
\\
&+&   \frac{1}{4} \sum_{ij}
\left[ \left( D_{ij}^{xx} - 2iD_{ij}^{xy} - D_{ij}^{yy} \right)  b_i^\dagger b_i b_i b_j + \textrm{h.c.} \right],
 \nonumber
 \\
 & &
\end{eqnarray}
where we have abbreviated $n_i = b^{\dagger}_i b_i$.
Next, we Fourier transform the Hamiltonian to momentum space,
setting
\begin{equation}
b_i =  \frac{1}{\sqrt{N}} \sum_{\bd{k}} e^{ i \bd{k} \cdot \bd{r}_i } b_{\bd{k}},
\label{eq:bFT}
\end{equation}
where for simplicity we impose periodic boundary conditions in all directions.
A more accurate calculation should take into account the finite extend in the direction
where the experimentally relevant samples have the smallest extension 
(which we call the $x$-direction)~\cite{Kreisel09}.
For our purpose it is sufficient to impose periodic boundary conditions in all directions,
which amounts to approximating the eigenfunctions of the exchange matrix $J_{ij}$
by plane waves. The lowest magnon band
is then obtained by simply setting  $k_x=0$.
In Ref.~\cite{Kreisel09}
we have shown that this {\it{uniform mode approximation}} reproduces the qualitative features of the dispersion of the lowest magnon mode rather well.
In momentum space, the quadratic part $H_2 ( t )$ of our bosonized Hamiltonian becomes
\begin{eqnarray}
{H}_2 &=&  \sum_{\bd k} \left[  A_{\bd k} b_{\bd k}^\dagger   b_{\bd k} +  
 \frac{B_{\bd k}}{2} \left(   b_{\bd k}^{\dagger}   b^{\dagger}_{-\bd k} +   b_{- \bd k}  
b_{\bd k}  \right) \right]\nonumber \\
&& + h_{1}\cos(\omega_{0}t)\sum_{\bd k}b_{\bd k}^{\dagger}b_{\bd k}, 
\end{eqnarray}
with
\begin{eqnarray}
A_{\bd k} & = & \sum_{i}e^{-i\bd k\cdot\bd r_{ij}}A_{ij}, \; \; \; 
B_{\bd k}  =  \sum_{i}e^{-i\bd k\cdot\bd r_{ij}}B_{ij}.
 \hspace{7mm}
\end{eqnarray}
The interaction parts can be written as
\begin{eqnarray}
 {H}_{3} & = & \frac{1}{\sqrt{N}}\sum_{\bd k_{1}\bd k_{2}\bd k_{3}}
\delta_{\bd k_{1}+\bd k_{2}+\bd k_{3} ,0}  \nonumber \\
&& \times
\frac{1}{2!} \Bigl[
 \Gamma_{1;23}^{\bar{b}bb}b_{-1}^{\dagger}b_{2}b_{3}    +\Gamma_{12;3}^{\bar{b}\bar{b}b}b_{-1}^{\dagger}b_{-2}^{\dagger}b_{3}
\Bigr], 
 \label{eq:H3hp}
\\
{H}_{4} & = & \frac{1}{N}\sum_{\bd k_{1}\dots\bd k_{4}}
\delta_{\bd k_{1}+\bd k_{2}+\bd k_{3}+\bd k_{4},0}
\Bigl[ \frac{1}{(2!)^{2}}\Gamma_{12;34}^{\bar{b}\bar{b}bb}b_{-1}^{\dagger}b_{-2}^{\dagger}b_{3}b_{4} \nonumber \\
 &  & \hspace{-7mm} + \frac{1}{3!}\Gamma_{1;234}^{\bar{b}bbb}b_{-1}^{\dagger}b_{2}b_{3}b_{4} 
+\frac{1}{3!}\Gamma_{123;4}^{\bar{b}\bar{b}\bar{b}b}b_{-1}^{\dagger}b_{-2}^{\dagger}b_{-3}^{\dagger}b_{4} \Bigr],
 \label{eq:H4hp}
\end{eqnarray}
where the properly symmetrized three-point  vertices are 
 \begin{eqnarray}
 \Gamma_{1;23}^{\bar{b}bb} & = & \sqrt{\frac{S}{2}}
\Big[ D_{\bd{k}_2 }^{zy}       -iD_{ \bd{k}_2 }^{zx}    + ( \bd{k}_2 \rightarrow \bd{k}_3 ) 
 \nonumber
 \\
 & & \hspace{8mm}
 + \frac{1}{2}(D_{ \bd{0} }^{zy}-iD_{ \bd{0} }^{zx})
   \Big],
 \label{eq:Gammabbb1}
 \\
 \Gamma_{12;3}^{\bar{b}\bar{b}b} & = & \big(\Gamma_{3;21}^{\bar{b}bb}\big)^{\ast},
 \label{eq:Gammabbb2}
\end{eqnarray}
and the symmetrized four-point vertices are
\begin{eqnarray}
\Gamma_{12;34}^{\bar{b}\bar{b}bb} & = & 
-  \frac{1}{2} \Bigl[     
  J_{\bd{k}_1 + \bd{k}_3 } +   J_{\bd{k}_2 + \bd{k}_3 }
  + J_{\bd{k}_1 + \bd{k}_4 } + J_{\bd{k}_2 + \bd{k}_4 }
 \nonumber
 \\
 & & \hspace{5mm} +
D_{ \bd{k}_1 + \bd{k}_3}^{zz} +
D_{ \bd{k}_2 + \bd{k}_3}^{zz} +
D_{ \bd{k}_1 + \bd{k}_4}^{zz} +
D_{ \bd{k}_2 + \bd{k}_4}^{zz} 
 \nonumber
 \\
 & & \hspace{5mm}
 - \sum_{i =1}^4 
 ( J_{\bd{k}_i } 
-  2    D_{ \bd{k}_i }^{zz} )
 \Bigr] , 
 \label{eq:UU1}
\\
 \Gamma_{1;234}^{\bar{b}bbb} & = & \frac{1}{4}
\Bigl[ D_{\bd{k}_2  }^{xx}-2iD_{ \bd{k}_2 }^{xy}-D_{\bd{k}_2}^{yy}
 \nonumber
 \\
& & \hspace{2mm}
+  ( \bd{k}_2 \rightarrow \bd{k}_3 ) + ( \bd{k}_2 \rightarrow \bd{k}_4 ) 
 \Big],
 \label{eq:UU2}
 \\
\Gamma_{123;4}^{\bar{b}\bar{b}\bar{b}b} & =&  \big( \Gamma_{4;123}^{\bar{b}bbb} \big)^\ast.
 \label{eq:UU3}
\end{eqnarray}
The Fourier transforms of the exchange and dipolar couplings are defined by
 \begin{eqnarray}
 J_{\bd{k}} & = & \sum_i e^{ - i \bd{k} \cdot \bd{r}_{ij} } J_{ ij },
 \\
 D_{\bd{k}}^{\alpha \beta} & = & \sum_i e^{ - i \bd{k} \cdot \bd{r}_{ij} }
D^{\alpha \beta}_{ ij }.
\end{eqnarray}
Finally, we use a Bogoliubov transformation to diagonalize the
time-independent part of ${H}_2 ( t )$,
\begin{eqnarray}
 \left( \begin{array}{c}
  b_{\bd{k}} \\
 b^{\dagger}_{-\bd{k}}
 \end{array}
 \right) = \left( \begin{array}{cc} u_{\bd{k}} & - v_{\bd{k}} \\
 -  v_{\bd{k}}^{\ast} & u_{\bd{k}}  \end{array} \right) 
\left( \begin{array}{c}
  a_{\bd{k}} \\
 a^{\dagger}_{-\bd{k}}
 \end{array}
 \right),
 \label{eq:Bogoliubov}
\end{eqnarray}
where
 \begin{equation}
 u_{\bd{k}} = \sqrt{ \frac{ A_{\bd{k}} + \epsilon_{\bd{k}} }{ 2 \epsilon_{\bd{k}} } } 
 \; \;   , \; \;
 v_{\bd{k}} = \frac{ B_{\bd{k}}}{ | B_{\bd{k}}    | }
\sqrt{ \frac{ A_{\bd{k}} - \epsilon_{\bd{k}} }{ 2 \epsilon_{\bd{k}} } } ,
 \end{equation}
and
 \begin{equation}
 \epsilon_{\bd{k}} = \sqrt{ A_{\bd{k}}^2 - | B_{\bd{k}} |^2 }.
 \end{equation}
After this transformation the quadratic part of the Hamiltonian reads \cite{Kloss10,Lvov94}
 \begin{eqnarray}
{H}_2 ( t ) &= & \sum_{\bd{k}} 
 \left[ \epsilon_{\bd{k}} a^{\dagger}_{\bd{k}} a_{\bd{k}}  + \frac{ \epsilon_{\bd{k}}
 - A_{\bd{k}} }{2} \right]
 \hspace{30mm}
 \nonumber
 \\
 &  & + h_1 \cos ( \omega_0 t )  \sum_{\bd{k}} 
 \left[ \frac{A_{\bd{k}}}{\epsilon_{\bd{k}}} a^{\dagger}_{\bd{k}} a_{\bd{k}} +   \frac{A_{\bd{k}} - \epsilon_{\bd{k}} }{
2 \epsilon_{\bd{k}} } \right]
\nonumber
 \\
& &  + \cos ( \omega_0 t )\sum_{\bd{k}} \left[ 
 \gamma_{\bd{k}} 
  a^{\dagger}_{\bd{k}} a^{\dagger}_{-\bd{k}} 
+
\gamma_{\bd{k}}^{\ast}
a_{-\bd{k}} a_{\bd{k}}
 \right],
 \label{eq:Hparpump}
 \end{eqnarray}
where
 \begin{equation}
 \gamma_{\bd{k}} = - \frac{h_1 B_{\bd{k}}}{ 2 \epsilon_{\bd{k}}} .
 \end{equation}
Substituting the Bogoliubov transformation (\ref{eq:Bogoliubov})
into the expressions for $H_3$ and $H_4$ given in
Eqs.~(\ref{eq:H3hp}, \ref{eq:H4hp}), we arrive at
expressions given in Eqs.~(\ref{eq:H3def}, \ref{eq:H4def}),
with the cubic vertices explicitly given by
\begin{subequations} 
\begin{eqnarray}
\Gamma_{123}^{aaa} & = & -\Gamma_{1;23}^{\bar{b}bb}v_{1}u_{2}u_{3}-\Gamma_{2;13}^{\bar{b}bb}v_{2}u_{1}u_{3}-\Gamma_{3;12}^{\bar{b}bb}v_{3}u_{1}u_{2}
 \nonumber \\
& & +\Gamma_{12;3}^{\bar{b}\bar{b}b}v_{1}v_{2}u_{3}+\Gamma_{23;1}^{\bar{b}\bar{b}b}v_{2}v_{3}u_{1}+\Gamma_{13;2}^{\bar{b}\bar{b}b}v_{1}v_{3}u_{2},\nonumber \\ 
& & \label{eq:Gammaaaa1}
\\
\Gamma^{\bar{a}aa}_{1;2,3} & = & \Gamma_{1;23}^{\bar{b}bb}u_{1}u_{2}u_{3}+\Gamma_{2;13}^{\bar{b}bb}v_{1}v_{2}u_{3}+\Gamma_{3;12}^{\bar{b}bb}v_{1}v_{3}u_{2} \nonumber \\
&& -\Gamma_{32;1}^{\bar{b}\bar{b}b}v_{3}v_{2}v_{1}-\Gamma_{12;3}^{\bar{b}\bar{b}b}v_{2}u_{1}u_{3}-\Gamma_{13;2}^{\bar{b}\bar{b}b}v_{3}u_{1}u_{2},\nonumber \\
& & \label{Gammaaaa2}
\\
 \Gamma_{12;3}^{\bar{a}\bar{a}a} & = & \big( \Gamma_{3;21}^{\bar{a}aa}\big)^{\ast},
 \label{eq:Gammaaaa3}
 \\
\Gamma_{123}^{\bar{a}\bar{a}\bar{a}} & = & \big(\Gamma_{123}^{aaa}\big)^{\ast}.
 \label{eq:Gammaaaa4}
 \end{eqnarray}
 \end{subequations} 
The vertices appearing in the quartic part $H_4$ of the Hamiltonian
in the Bogoliubov basis are (see Eq.~(\ref{eq:H4def}))
\begin{subequations} 
\begin{eqnarray}
\Gamma_{1234}^{aaaa} & = & 
\Gamma_{12;34}^{\bar{b}\bar{b}bb}u_{1}u_{2}v_{3}v_{4}+\Gamma_{13;24}^{\bar{b}\bar{b}bb}u_{1}u_{3}v_{2}v_{4}
\notag\\&&\hspace{-1cm}
+\Gamma_{14;23}^{\bar{b}\bar{b}bb}u_{1}u_{4}v_{2}v_{3}+\Gamma_{23;14}^{\bar{b}\bar{b}bb}u_{2}u_{3}v_{1}v_{4}+\Gamma_{24;13}^{\bar{b}\bar{b}bb}u_{2}u_{4}v_{1}v_{3}
\notag\\&&\hspace{-1cm}
+\Gamma_{34;12}^{\bar{b}\bar{b}bb}u_{3}u_{4}v_{1}v_{2}
 -\Gamma_{4;123}^{\bar{b}bbb}u_{1}u_{2}u_{3}v_{4}-\Gamma_{3;124}^{\bar{b}bbb}u_{1}u_{2}u_{4}v_{3}
\notag\\&&\hspace{-1cm}
-\Gamma_{2;134}^{\bar{b}bbb}u_{1}u_{3}u_{4}v_{2}-\Gamma_{1;234}^{\bar{b}bbb}u_{2}u_{3}u_{4}v_{1}
  -\Gamma_{234;1}^{\bar{b}\bar{b}\bar{b}b}u_{1}v_{2}v_{3}v_{4}
\notag\\&&\hspace{-1cm}
-\Gamma_{134;2}^{\bar{b}\bar{b}\bar{b}b}u_{2}v_{1}v_{3}v_{4}-\ \Gamma_{124;3}^{\bar{b}\bar{b}\bar{b}b}u_{3}v_{1}v_{2}v_{4}-\Gamma_{123;4}^{\bar{b}\bar{b}\bar{b}b}u_{4}v_{1}v_{2}v_{3},
\label{eq:Uaa1}
\notag\\
\\
\Gamma_{1;234}^{\bar{a}aaa} & = & 
-\Gamma_{21;34}^{\bar{b}\bar{b}bb}u_{2}v_{1}v_{3}v_{4}-\Gamma_{31;24}^{\bar{b}\bar{b}bb}u_{3}v_{1}v_{2}v_{4}
\notag\\&&\hspace{-1cm}
-\Gamma_{41;23}^{\bar{b}\bar{b}bb}u_{4}v_{1}v_{2}v_{3}-\Gamma_{34;12}^{\bar{b}\bar{b}bb}u_{3}u_{4}u_{1}v_{2}-\Gamma_{24;13}^{\bar{b}\bar{b}bb}u_{2}u_{4}u_{1}v_{3}
\notag\\&&\hspace{-1cm}
-\Gamma_{23;14}^{\bar{b}\bar{b}bb}u_{2}u_{3}u_{1}v_{4}
  +  \Gamma_{1;234}^{\bar{b}bbb}u_{1}u_{2}u_{3}u_{4}+\Gamma_{4;321}^{\bar{b}bbb}u_{3}u_{2}v_{1}v_{4}
\notag\\&&\hspace{-1cm}
+\Gamma_{3;421}^{\bar{b}bbb}u_{4}u_{2}v_{1}v_{3}+\Gamma_{2;431}^{\bar{b}bbb}u_{4}u_{3}v_{1}v_{2}
  +  \Gamma_{123;4}^{\bar{b}\bar{b}\bar{b}b}u_{4}u_{1}v_{2}v_{3}
\notag\\&&\hspace{-1cm}
+\Gamma_{124;3}^{\bar{b}\bar{b}\bar{b}b}u_{3}u_{1}v_{2}v_{4}+\Gamma_{134;2}^{\bar{b}\bar{b}\bar{b}b}u_{2}u_{1}v_{3}v_{4}+\Gamma_{432;1}^{\bar{b}\bar{b}\bar{b}b}v_{4}v_{2}v_{3}v_{1},
\label{eq:Uaa2}
\notag\\
\\
\Gamma_{12;34}^{\bar{a}\bar{a}aa} & = & 
\Gamma_{12;34}^{\bar{b}\bar{b}bb}u_{1}u_{2}u_{3}u_{4}+\Gamma_{13;42}^{\bar{b}\bar{b}bb}u_{1}u_{4}v_{3}v_{2}
\notag\\&&\hspace{-1cm}
+\Gamma_{14;32}^{\bar{b}\bar{b}bb}u_{1}u_{3}v_{4}v_{2}+\Gamma_{23;41}^{\bar{b}\bar{b}bb}u_{2}u_{4}v_{3}v_{1}+\Gamma_{24;31}^{\bar{b}\bar{b}bb}u_{2}u_{3}v_{4}v_{1}
\notag\\&&\hspace{-1cm}
+\Gamma_{12;34}^{\bar{b}\bar{b}bb}v_{1}v_{2}v_{3}v_{4}
- \Gamma_{4;321}^{\bar{b}bbb}u_{3}v_{2}v_{1}v_{4}-\Gamma_{3;421}^{\bar{b}bbb}u_{4}v_{2}v_{1}v_{3}
\notag\\&&\hspace{-1cm}
-\Gamma_{2;341}^{\bar{b}bbb}u_{2}u_{3}u_{4}v_{1}-\Gamma_{1;342}^{\bar{b}bbb}u_{1}u_{3}u_{4}v_{2}
- \Gamma_{234;1}^{\bar{b}\bar{b}\bar{b}b}u_{2}v_{3}v_{4}v_{1}
\notag\\&&\hspace{-1cm}
-\Gamma_{134;2}^{\bar{b}\bar{b}\bar{b}b}u_{1}v_{3}v_{4}v_{2}-\Gamma_{124;3}^{\bar{b}\bar{b}\bar{b}b}u_{1}u_{2}u_{3}v_{4}-\Gamma_{123;4}^{\bar{b}\bar{b}\bar{b}b}u_{1}u_{2}u_{4}v_{3},
 \label{eq:Uaa3}
\notag\\
\end{eqnarray}
and
 \begin{eqnarray}
 \Gamma_{1234}^{aaaa} & = & \Gamma_{1234}^{\bar{a}\bar{a}\bar{a}\bar{a}},
 \label{eq:Uaa4}
 \\
 \Gamma_{4;321}^{\bar{a}aaa} & = &  
\big(\Gamma_{123;4}^{\bar{a}\bar{a}\bar{a}a} \big)^\ast. \label{eq:Uaa5}
 \end{eqnarray}
\end{subequations}
For nearest neighbor coupling on a cubic lattice with spacing $a$ the
Fourier transform of the exchange coupling  appearing in the above expressions is
\begin{equation}
J_{{\bd k}} = 2 J [  \cos ( k_x a ) + \cos(k_y a) +  \cos(k_z a)  \big].
\end{equation}
The Fourier transform of the dipolar tensor is more complicated. 
For a thin YIG film the minimum of the
dispersion is at $\pm \bd{q} = ( 0,0,  \pm q )$, so that in this work
we only need $D^{\alpha \beta}_{ \bd{k}}$ as a function of $k_z$ for
 $k_x = k_y= 0$. 
For a  film with thickness $d \gg a$ we then obtain in  uniform mode 
approximation~\cite{Kreisel09},
\begin{subequations}
\begin{eqnarray}
D_{{ k}_z}^{xx} &=& \frac{4\pi \mu^2}{a^3} \left[ \frac{1}{3} - f_{ k_z} \right], \\
D_{k_z}^{yy} &=& \frac{4\pi \mu^2}{ 3 a^3} , \\
D_{{ k}_z}^{zz} &=& \frac{4\pi \mu^2}{a^3} \left[ -\frac{2}{3} 
+  f_{{ k}_z}  \right], \\
D_{{k}_z}^{xy} &=&  D_{{ k}_z}^{xz} = D_{k_z}^{yz}  = 0 ,
\end{eqnarray}
\end{subequations}
where 
the form factor $f_{{k}_z}$ is given by~\cite{Kalinikos86,Kreisel09}
 \begin{equation}
f_{k_z}=\frac{1-e^{-|k_z |d}}{|{ k}_z|d}.
 \end{equation}

For the purpose of studying 
the phenomenon of parametric 
resonance~\cite{Rezende09,Lvov94,Zakharov70,Tsukernik75,Vinikovetskii79,Lim88,Kalafati89},
one usually simplifies the Hamiltonian (\ref{eq:Hparpump})  by  dropping
the second line.
involving the combination $\cos ( \omega_0 t ) A_{\bd{k}} a^{\dagger}_{\bd{k}} a_{\bd{k}}$;
moreover in the last line one substitutes
 \begin{equation}
 \gamma_{\bd{k}} \cos ( \omega_0 t ) \rightarrow 
 \frac{\gamma_{\bd{k}}}{2} e^{ - i \omega_0 t } \; \; , 
 \; \;
\gamma_{\bd{k}}^{\ast} \cos ( \omega_0 t ) \rightarrow 
 \frac{\gamma_{\bd{k}}^{\ast} }{2} e^{  i \omega_0 t }.
 \label{eq:resonanceapprox}
 \end{equation}
Although the
validity of this approximation in the context of YIG is
questionable \cite{Kloss10} (see also Ref.~\cite{Zvyagin82}), let us 
assume here that it correcly describes at least some aspects of 
the experiments~\cite{Demokritov06,Demidov07,Dzyapko07,Demidov08,Demokritov08}.
Our quadratic boson Hamiltonian is then approximated by
\begin{eqnarray}
 {H}_{ 2} ( t ) &  \approx &  
\sum_{\bd{k}} \epsilon_{\bd{k}} a_{\bd{k}}^{\dagger} a_{\bd{k}} 
 \nonumber
 \\
& + &  
\frac{1}{2} \sum_{\bd{k}}
\left[    
\gamma_{\bd{k}} e^{ -i \omega_0 t }  a^{\dagger}_{\bd{k}} a^{\dagger}_{ - \bd{k}}
+ \gamma_{\bd{k}}^{\ast} e^{  i \omega_0 t }  a_{- \bd{k}} a_{\bd{k}}    
  \right],
 \label{eq:H2pr}
 \hspace{10mm}
 \end{eqnarray}
where we have dropped the constant terms.
The explicit time-dependence may now be removed by a canonical transformation to the rotating
reference frame,
 \begin{equation}
  \tilde{a}_{\bd{k}}  = e^{ i \omega_0 t /2 } a_{\bd{k}}, \; \; \; 
 \tilde{a}_{\bd{k}}^{\dagger}  = e^{ - i \omega_0 t /2 } a_{\bd{k}},
 \end{equation}
so that the transformed quadratic part of our Hamiltonian is
 \begin{equation}
 \tilde{H}_2 = \sum_{\bd{k}} 
 \left[ \tilde{\epsilon}_{\bd{k}}  \tilde{a}^{\dagger}_{\bd{k}} \tilde{a}_{\bd{k}}
 + \frac{\gamma_{\bd{k}}}{2} \tilde{a}^{\dagger}_{\bd{k}} \tilde{a}^{\dagger}_{ - \bd{k}} +
\frac{\gamma^{\ast}_{\bd{k}}}{2} \tilde{a}_{-\bd{k}} \tilde{a}_{  \bd{k}}  \right],
 \label{eq:H2defrot}
 \end{equation}
where $\tilde{\epsilon}_{\bd{k}} = \epsilon_{\bd{k}} - \omega_0 /2$.
If we re-define again $\tilde{a}_{\bd{k}} \rightarrow a_{\bd{k}}$, 
$\tilde{\epsilon}_{\bd{k}} \rightarrow \epsilon_{\bd{k}}$, we arrive at
the quadratic Hamiltonian in Eq.~(\ref{eq:H2def}).

\end{appendix}


\begin{thebibliography}{99}
%
\bibitem{Demokritov06}
S.~O. Demokritov, V.~E. Demidov, O.~Dzyapko, G.~A. Melkov, A.~A. Serga,
  B.~Hillebrands, and A.~N. Slavin, Nature {\bf 443}, 430 (2006).
%
\bibitem{Demidov07}
V.~E. Demidov, O.~Dzyapko, S.~O. Demokritov, G.~A. Melkov, and A.~N. Slavin,
  Phys. Rev. Lett. {\bf 99}, 037205 (2007).
%
\bibitem{Dzyapko07}
O.~Dzyapko, V.~E. Demidov, S.~O. Demokritov, G.~A. Melkov, and A.~N. Slavin,
  New J. Phys. {\bf 9}, 64 (2007).
%
\bibitem{Demidov08}
V.~E. Demidov, O.~Dzyapko, S.~O. Demokritov, G.~A. Melkov, and A.~N. Slavin,
  Phys. Rev. Lett. {\bf 100}, 047205 (2008).
%
\bibitem{Demokritov08}
S.~O. Demokritov, V.~E. Demidov, O.~Dzyapko, G.~A. Melkov, and A.~N. Slavin,
  New J. Phys. {\bf 10}, 045029 (2008).
%
\bibitem{Kalinikos86}
B. A. Kalinikos and A. N. Slavin, J. Phys. C {\bf{19}}, 7013 (1986);
J. Phys. Condens. Matter {\bf{2}}, 9861 (1990).
%
\bibitem{Kreisel09}
A.~Kreisel, F.~Sauli, L.~Bartosch, and P.~Kopietz,
\newblock Eur. Phys. J. B {\bf 71}, 59 (2009).
%
\bibitem{Ueda09}
H. T. Ueda and K. Totsuka, Phys. Rev. B {\bf{80}}, 014417 (2009).
%
\bibitem{Yukalov78}
V. I. Yukalov, Teor. Mat. Fiz. {\bf{37}}, 390 (1978)
[Theoret. and Math. Phys. {\bf{37}}, 1093 (1978)].
%
\bibitem{Alexander78}
S.  Alexander and J. P. McTague, Phys. Rev. Lett. {\bf{41}}, 702  (1978).
%
\bibitem{Anderson84}
The Landau theory of the liquid-solid transition has been discussed by
P. W. Anderson in {\it{Basic Notions of Condensed Matter Physics}}
(Benjamin/Cummings, Menlo Park, CA, 1984); see also
P. M. Chaikin and T. C. Lubensky in {\it{Principles of condensed matter physics}}
(Cambridge University Press, Cambridge, 1995).
%
\bibitem{Kohn70}
W. Kohn and D. Sherrrington, Rev. Mod. Phys. {\bf{42}}, 1 (1970).
%
\bibitem{superfluidity}
Under equilibrium conditions the frictionless flow in a superfluid is only 
possible for flow speeds below the Landau critical velocity $v_c$. The latter is 
determined by the dispersion of the Bogoliubov quasi particle, which is the
gapless Goldstone mode associated with spontaneous breaking of the 
$U\left(1\right)$-symmetry in the superfluid state. However, the Hamiltonian 
for magnons in YIG in the parallel pumping geometry does not have $U\left(1\right)$-symmetry, 
so that there is no gapless Bogoliubov mode. The Landau criterion of 
superfluidity is therefore not applicable for magnons in YIG. On the other hand, 
even in the absence of a Landau critical velocity there can be frictionless 
transport of quasi particles under non-equilibrium condition, see M. Wouters and 
I. Carusotto, Phys. Rev. Lett. {\bf{105}}, 020602 (2010).
%
\bibitem{Brazovskii75}
S. A. Brazovskii, Zh. Eksp. Teor. Fiz. {\bf{68}}, 175 (1975)
[Sov. Phys. JETP {\bf{41}}, 85 (1975)].
%
\bibitem{Hohenberg95}
P. C. Hohenberg and J. B. Swift,  Phys. Rev. E {\bf{52}}, 1828 (1995). 
%
\bibitem{Tupitsyn08}
I. S. Tupitsyn, P. C. E. Stamp, and A. L. Burin,
Phys. Rev. Lett. {\bf{100}}, 257202 (2008).
%
\bibitem{Kloss10}
T. Kloss, A. Kreisel, and P. Kopietz, Phys. Rev. B {\bf{81}}, 104308 (2010).
%
\bibitem{Pitaevskii03}
See, for example, L. Pitaevskii and S. Stringari, {\it{Bose-Einstein Condensation}}
(Clarendon Press, Oxford, 2003).
%
\bibitem{Schuetz05}
F. Sch\"{u}tz, L. Bartosch, and P. Kopietz, Phys. Rev. B {\bf{72}}, 035107 (2005).
%
\bibitem{Ma76}
See, for example, S. K. Ma, {\it{Modern Theory of Critical Phenomena}}
(Benjamin/Cummings, Reading, MA, 1976).
%
\bibitem{DellAmore09}
R. Dell'Amore, A. Schilling, and K. Kr\"{a}mer, Phys. Rev. B {\bf{79}}, 014438 (2009).
%
\bibitem{Cherepanov93}
V. Cherepanov, I. Kolokolov, and V. L'vov,
Phys. Rept. {\bf{229}}, 81 (1993).
%
\bibitem{Rezende06}
S.~M. Rezende, F.~M. de~Aguiar, and A.~Azevedo,
Phys. Rev. B {\bf 73}, 094402 (2006).
%
\bibitem{Rezende09}
S.~M. Rezende,
Phys. Rev. B {\bf 79}, 060410 (2009); 
{\it{ibid.}} {\bf{79}},  174411  (2009).
%
\bibitem{Holstein40}
T.~Holstein and H.~Primakoff,
\newblock Phys. Rev. {\bf 58}, 1098 (1940).
%
\bibitem{Lvov94}
V. S. L'vov, {\it{Wave Turbulence Under Parametric Excitations}}, (Springer, Berlin, 1994).
%
\bibitem{Zakharov70}
V. E. Zakharov, V. S. L'vov, and S. S. Starobinets,
Zh. Eksp. Teor. Fiz. {\bf{59}}, 1200 (1970) 
[Sov. Phys. JETP {\bf{32}}, 656 (1971)];
V. E. Zakharov, V. S. L'vov, and S. S. Starobinets,
Usp. Fiz. Nauk {\bf{114}}, 609 (1974) [Sov. Phys.-Usp. {\bf{17}}, 896 (1975)].
%
\bibitem{Tsukernik75}
V. M. Tsukernik and R. P. Yankelevich,
Zh. Eksp. Teor. Fiz. {\bf{68}}, 2116 (1975) [Sov. Phys. JETP {\bf{41}}, 1059 (1976)].
%
\bibitem{Vinikovetskii79}
I. A. Vinikovetskii, A. M. Frishman, and V. M. Tsukernik,
Zh. Eksp. Teor. Fiz. {\bf{76}}, 2110 (1979) [Sov. Phys. JETP {\bf{49}}, 1067 (1979)].
%
\bibitem{Lim88}
S. P. Lim and D. L. Huber, Phys. Rev. B {\bf{37}}, 5426 (1988);
{\it{ibid.}} {\bf{41}}, 9283 (1990).
%
\bibitem{Kalafati89}
Yu. D. Kalafati and V. L. Safanov,
Zh. Eksp. Teor. Fiz. {\bf{95}}, 2009 (1989) 
[Sov. Phys. JETP {\bf{68}}, 1162 (1989)].
%
\bibitem{Zvyagin82}
A. A. Zvyagin, V. Ya. Serebryannyi, A. M. Frishman, and V. M. Tsukernik,
Fiz. Nizk. Temp. {\bf{8}}, 1205 (1982) [Sov. J. Low Temp. Phys.
 {\bf{8}}, 612 (1982)].
%
\bibitem{Malomed10}
B. A. Malomed, O. Dzyapko, V. E. Demidov, and S. O. Demokritov,
Phys. Rev. B {\bf{81}}, 024418 (2010).
%
\end{thebibliography}
\end{document}